# Designing strong and tunable magnetoelectric coupling in 2D trilayer heterostructures

Xin Jin,[a] Andrew O' Hara,[b] Yu-Yang Zhang,[a] Shixuan Du,[a,c,*]

and Sokrates T. Pantelides[b,a,*]

[a] Institute of Physics & University of the Chinese Academy of Sciences, Chinese Academy of Sciences, Beijing 100190, China.

[b] Department of Physics and Astronomy and Department of Electrical and Computer Engineering, Vanderbilt University, Nashville, TN 37235, USA

[c] Songshan Lake Materials Laboratory, Dongguan, Guangdong 523808, China

*Corresponding authors: sxdu@iphy.ac.cn, pantelides@vanderbilt.edu

## Abstract

The quest for electric-field control of nanoscale magnetic states such as skyrmions, which would impact the field of spintronics, has led to a challenging search for multiferroic materials or structures with strong magnetoelectric coupling and efficient electric-field control. Here we report a theoretical prediction that such phenomena can be realized in two-dimensional (2D) bilayer FE/PMM and trilayer FE/PMM/FE heterostructures (two-terminal and three-terminal devices), where FE is a 2D ferroelectric and PMM is a polar magnetic metal with strong spin-orbit coupling. Such a PMM has strong Dzyaloshinskii-Moriya interactions (DMI) that can generate skyrmions, while the FE can generate strong magnetoelectric coupling through polarization-polarization interactions. In trilayer heterostructures, contact to the metallic PMM layer enables multiple polarization configurations for electric-field control of skyrmions. We report density-functional-theory calculations for particular material choices that demonstrate the effectiveness of these arrangements, with the key driver being the polarization-polarization interactions between the PMM and FE layers. The present findings provide a method to achieve strong magnetoelectric coupling in the 2D limit and a new perspective for the design of related spintronics.

Topologically protected, noncollinear nanoscale spin textures in magnetic materials, e.g., magnetic skyrmions, bimerons, etc., hold promise for applications in next-generation data-storage devices and spintronics[1-3]. The utility of such magnetic structures would be enhanced significantly if they can be generated or manipulated by applied electric fields. Thus, one must identify multiferroic materials or heterostructures that feature magnetic properties favoring noncollinear magnetism, ferroelectricity (FE), and strong magnetoelectric coupling, i.e., large $\partial P/\partial H$ or $\partial M/\partial E$.

In magnetic materials, the ratio of Dzyaloshinskii-Moriya interaction (DMI) $d$ and the exchange integral $J$ controls the magnetic ordering – large $J$ produces collinear ferromagnetism or antiferromagnetism, while large DMI produces noncollinear orderings. In order to have large DMI, it is necessary for the magnetic material to lack inversion symmetry and have strong spin-orbit coupling (SOC)[4,5]. Thus, magnetic skyrmions and bimerons are usually observed in magnets without inversion symmetry[6-10] and noncentrosymmetric thin-film heterostructures comprising a ferromagnetic (FM) metal film and one or more heavy-metal (HM) films, e.g., Pt/Co/MgO and Pt/Co/Ir[11-13].

It follows that a targeted optimal FM/FE multiferroic system must feature large DMI (noncentrosymmetry, large SOC) and strong magnetoelectric coupling. One would expect that all these conditions might be satisfied in a single FM/FE multiferroic material and magnetoelectric coupling would be strong because the FM/FE properties would be intimately linked. So far, control of polarization by magnetic field has been achieved in bulk multiferroics[14-16], however, control of magnetism by electric fields has not proved to be very practical in bulk multiferroics[17-19]. In 2018, magnetoelectric coupling was demonstrated in a FE/FM heterostructure of perovskite-structure transition-metal oxides, namely $BaTiO_3/SrRuO_3$ (BTO/SRO)[20]. Magnetoelectric coupling is achieved by electric-field-controlled polarization reversal in the FE material. We note, however, that the magnetic material, SRO, is centrosymmetric, whereby nonzero DMI is generated only by the breaking of inversion symmetry by the presence of BTO. As a result, the induced DMI values in SRO are quite small, ranging from zero to ~1.25 meV and from zero to ~0.6 meV in the SRO layer in the two polarization directions. Additionally, the corresponding small DMI changes induced by polarization reversal indicate relatively weak magnetoelectric coupling.

The recent discovery of two-dimensional (2D) FM[21-23] and FE[24-28] materials triggered interest in 2D FM/FE heterostructures for skyrmion generation and manipulation. Several theoretical papers have predicted the possibility of generating and annihilating skyrmions or bimerons in such 2D heterostructures[29-31]. Once more, however, we note that, as in the BTO/SRO thin-film heterostructure, the FM materials in the studied 2D heterostructures are centrosymmetric, resulting in DMI values that are < 1 meV in both polarization directions.

In the present work we overcome the above limitations and design optimal bilayer and trilayer heterostructures exhibiting strong and, in the trilayer, tunable magnetoelectric coupling for the manipulation of skyrmions and other noncollinear spin structures. We first identify 2D materials that satisfy the aforementioned criteria for large DMI [noncentrosymmetric (polar) FM 2D materials that also have strong SOC, i.e., contain heavy elements], which are needed for the generation of stable noncollinear spin structures. We then combine a *polar* 2D FM material and a 2D FE material to construct 2D FM/FE bilayer and FE/FM/FE trilayer heterostructures. In such structures, strong magnetoelectric coupling would arise via polarization-polarization interactions between the polar-FM and FE materials, causing large changes of magnetic parameters, i.e., $\Delta d$, $\Delta J$, etc., during FE polarization reversal. Additionally, for the 2D FM material, a metallic one is more desirable as it can function as an electrode in a FE/FM/FE trilayer heterostructure, enabling multiple polarization configurations, i.e., tuning the magnetoelectric strength. Polar metals were theoretically proposed in 1965 by Anderson and Blount[33], but the first realization of polar metals was reported in 2013[32], and they remain relatively rare. Magnetic 2D materials were first reported in 2017 and also remain relatively rare[21-23]. In 2020, two 2D materials were reported that meet the desired criteria: The Janus transition-metal dichalcogenides MnSTe and MnSeTe are stable metallic polar 2D magnetic metals with large SOC and, as expected, they have been predicted to have large DMI values and skyrmions[34,35]. We, therefore, propose bilayer MnSeTe/In$_2$Se$_3$ heterostructures, where monolayer In$_2$Se$_3$ is the 2D FE material, and trilayer In$_2$Se$_3$/MnSeTe/In$_2$Se$_3$ heterostructures, which can form the core of a three-terminal device. MnSeTe is selected to construct heterostructures because MnSe$_2$ has been fabricated[36], whereby the Janus MnSeTe is likely to be fabricated. Single-phase FE In$_2$Se$_3$ has been fabricated in multilayer form[25,37,38]. We perform calculations using a single quintuple layer In$_2$Se$_3$ for computational convenience, but multilayers would work just as well, or perhaps even better by providing tunability of the polarization in the FE layer in heterostructure. We employ DFT calculations to demonstrate that strong magnetoelectric coupling is indeed achieved in the proposed heterostructures. In the trilayer In$_2$Se$_3$/MnSeTe/In$_2$Se$_3$ heterostructure, we also find that through the combination of multiple polarization states, both tunable magnetoelectric coupling and control of the magnetic-skyrmion evolution in MnSeTe can be achieved. The present findings provide a method to achieve strong magnetoelectric coupling in the 2D limit, as well as a new perspective for the design of related spintronics.

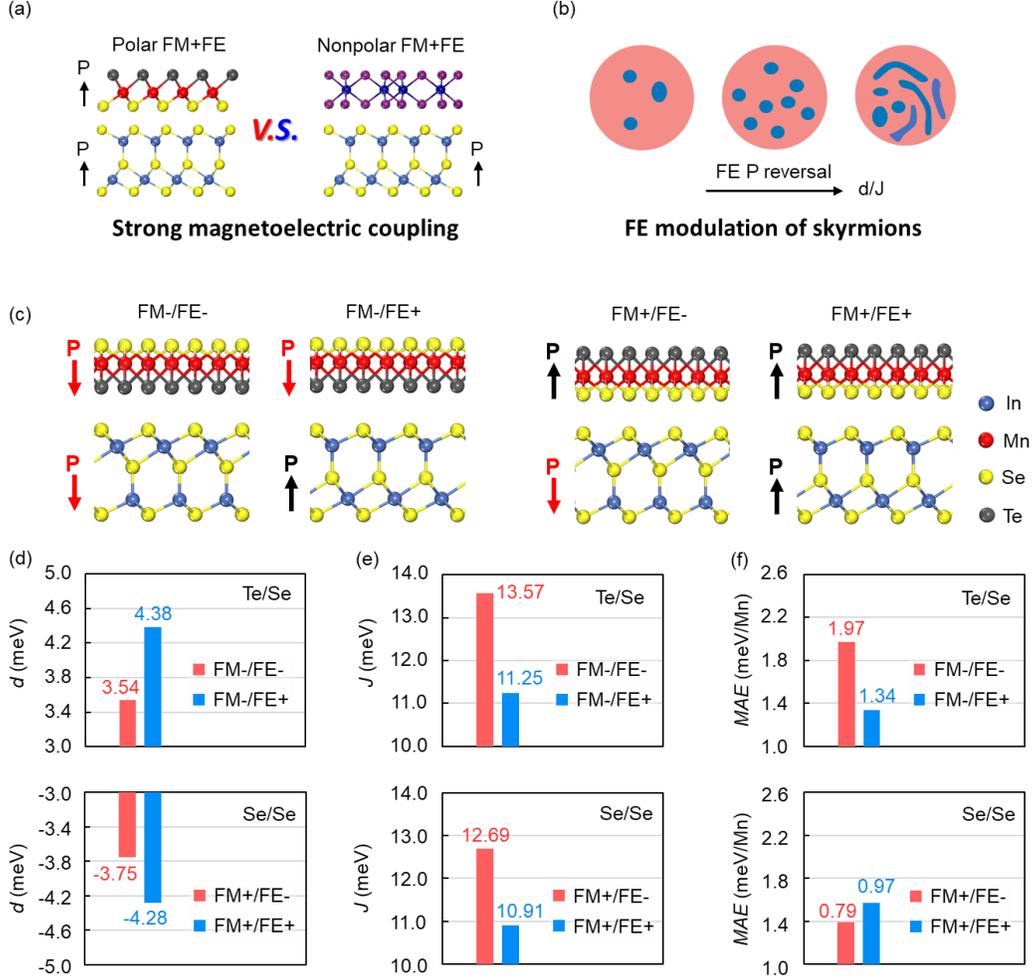

**Figure 1. Magnetoelectric coupling in MnSeTe/In$_2$Se$_3$ heterostructure.** Schematic illustration of (a) a polar-FM/FE heterostructure (left), in which there exists a strong polarization-polarization interaction, and a nonpolar-FM/FE heterostructure (right), in which the polarization of In$_2$Se$_3$ interacts only with very small induced polarization in the nonpolar-FM layer; and (b) the effect of polarization reversal in the In$_2$Se$_3$ layer on a skyrmion in the polar-FM layer. (c) Possible configurations of the MnSeTe/In$_2$Se$_3$ heterostructure. The – and + signs represent the polarization directions. (d)-(f) The DMI parameter $d$, exchange parameter $J$, and magnetic-anisotropy energy ($MAE$) of MnSeTe in the heterostructure. Te/Se and Se/Se represent the interface of the heterostructure.

We first describe briefly how to characterize the expected properties exhibited in a MnSeTe/In$_2$Se$_3$ heterostructure, i.e., strong magnetoelectric coupling and FE polarization-controlled skyrmion evolution, through computational approaches. To demonstrate the strong magnetoelectric coupling in such heterostructures, we compare the polarization-reversal-induced changes of the magnetic parameters, including changes of the DMI parameter $d$, Heisenberg isotropic exchange parameters $J$, and magnetic-anisotropy energy ($MAE$), to the corresponding changes in a bilayer heterostructure composed of FE and nonpolar-FM 2D materials, as shown schematically in Fig. 1(a). Larger changes of magnetic parameters are expected in MnSeTe/In$_2$Se$_3$ heterostructure if magnetoelectric coupling in MnSeTe/In$_2$Se$_3$ is stronger. Additionally, the state of skyrmions, including the formation of skyrmions and skyrmion densities, can be described by the ratio between DMI and exchange parameters $d/J$[39]. Thus, as shown in Fig. 1(b), the FE polarization-controlled skyrmion evolution can be characterized by comparing the $d/J$ before and after polarization reversal. Other magnetic interactions, e.g., anisotropic exchange interaction, are not discussed here since the such interactions in MnSeTe are one order of magnitude smaller than DMI and $J$[34,35].

Given the above framework, we calculated the magnetic parameters of MnSeTe in a MnSeTe/In$_2$Se$_3$ heterostructure to check whether interlayer magnetoelectric coupling is established between MnSeTe and In$_2$Se$_3$. Considering the combinations of the polarizations of MnSeTe and In$_2$Se$_3$, there are total of four possible atomic configurations of MnSeTe/In$_2$Se$_3$ heterostructure classified into 2 types: one possesses a Te/Se interface and the other possesses a Se/Se interface, as shown in Fig. 1(c). These four configurations are named FM-/FE-, FM-/FE+, FM+/FE- and FM+/FE+, in which the signs – and + denote the polarization directions. The calculated magnetic parameters $d$, $J$ and $MAE$ of MnSeTe in the two distinct MnSeTe/In$_2$Se$_3$ heterostructures are shown in Figs. 1(d)-(f). Evidently, the strengths of $d$, $J$, and $MAE$ are modulated when reversing the FE polarization of the In$_2$Se$_3$ layer (from red bars to blue bars or the reverse), proving the existence of the interlayer magnetoelectric coupling in the MnSeTe/In$_2$Se$_3$ heterostructure.

To examine the strength of the magnetoelectric coupling, we further compare the modulation amplitudes of $d$, $J$ and $MAE$ of MnSeTe/In$_2$Se$_3$ to the corresponding values of some reported nonpolar-FM/FE systems, as listed in Table 1. The MnSeTe/In$_2$Se$_3$ heterostructures show significantly enhanced modulation of the magnetic parameters. For example, in a MnSeTe/In$_2$Se$_3$ heterostructure with Te/Se interface, the modulation amplitude of DMI parameters, $|\Delta d|$, is nearly six times larger than that of the Cr$_2$Ge$_2$Te$_6$/In$_2$Se$_3$ heterostructure, as summarized in Table 1, indicating that the magnetoelectric coupling in the MnSeTe/In$_2$Se$_3$ heterostructure is significantly stronger. The key difference between the present and the reported system is that, the FM layer in our system, MnSeTe, is polar. Thus, the stronger magnetoelectric coupling could be attributed to the polarization-polarization interactions between MnSeTe and In$_2$Se$_3$.

**Table 1. Comparison between the changes of DMI parameters $d$, exchange parameters $J$, magnetic-anisotropy energy $MAE$ and $d/J$ ratio of the present system and those in reported nonpolar-FM/FE heterostructures when the polarization of the FE layer is reversed. In the present results for the trilayer heterostructure, we list only the largest change values of $d$, $J$ and $MAE$ among all possible transitions (see Fig. 2). Arrows represent the FE polarization directions.**

|  | Present results | | MnBi$_2$Te$_4$/In$_2$Se$_3$[40] | Cr$_2$Ge$_2$Te$_6$/In$_2$Se$_3$[29,41] | LaCl/In$_2$Se$_3$[30] | | WTe$_2$/CrCl$_3$/CuInP$_2$S$_6$[31] |
|---|---|---|---|---|---|---|---|
|  | MnSeTe/In$_2$Se$_3$ | In$_2$Se$_3$/MnSeTe/In$_2$Se$_3$ |  |  |  |  |  |
| $|\Delta d|$ (meV) | 0.84 (Te/Se) 0.53 (Se/Se) | 1.25 (↑↑↓ to ↓↑↑) | No report | 0.14 | $\Delta d_1$ $\Delta d_2$ $\Delta d_3$ | 0.35 0.36 0.53 | 0.04 |
| $|\Delta J|$ (meV) | 2.32 (Te/Se) 1.78 (Se/Se) | 3.01 (↑↑↓ to ↓↑↑) | ~0.6 | 0.62 | $\Delta J_1$ $\Delta J_2$ $\Delta J_3$ | 1.94 2.39 3.36 | 1.81 |
| $|\Delta MAE|$ (meV) | 0.63 (Te/Se) 0.18 (Se/Se) | 0.66 (↑↑↓ to ↓↑↑) | ~0.1 | 0.17 | No report | | 0.13 |
| $d/J$ | 0.39 (↑, Te/Se) 0.26 (↓, Te/Se) | 0.22 (↑↑↑) 0.17 (↑↑↓) 0.31 (↓↑↑) 0.25 (↓↑↓) | No report | 0.017 (↑) 0.023 (↓) | $d_1/J_1$ $d_2/J_2$ $d_3/J_3$ | 0.43 (↑) 0.07 (↓) 0.09 (↑) 0.07 (↓) 0.10 (↑) 0.06 (↓) | 0.07 (↑) 0.04 (↓) |

Given that the emergence and evolution of magnetic skyrmions mainly relies on the competition between the DMI, which tilt spins, and exchange interactions, which align spins, the enhanced $d$ and suppressed $J$ during FE polarization reversal (as shown in Figs. 1(d)-(e), red bars to blue bars) indicate the possibility of controlling skyrmions in the MnSeTe/In$_2$Se$_3$ heterostructure, i.e., controlling the density or the creation (elimination) of skyrmions. Here we note that, for a MnSeTe/In$_2$Se$_3$ heterostructure with Se/Se interface, the DMI enhancement refers to the absolute values of $d$ getting larger, since the sign of $d$ represents only the spin whirling direction. We note that, as shown in Figs. 1(d)-(f), the values of $MAE$ are quite small relative to the values of $J$ and $d$, which are the main controllers of magnetic ordering. Furthermore, the value of $MAE$ is to be compared with the value of $d^2$ to consider the effect of $MAE$ on the formation of skyrmions[1]. Overall, the $MAE$ values are quite small and do not play a significant role in the processes of interest.

To further characterize the evolution of skyrmions, we calculate the $d/J$ ratio, as summarized in Table 1. The $d/J$ ratios of the MnSeTe/In$_2$Se$_3$ heterostructure are in the typical range for skyrmion formation (~0.1-0.3)[39] and comparable to that of freestanding MnSeTe monolayer (0.34)[35], indicating skyrmions also exist in the heterostructure. Comparing to the reported nonpolar-FM/FE heterostructure[29-31,40,41], the present system exhibits both large $d/J$ and $d$ as shown in Table 1 and Table S4, which benefits the formation and stabilization of skyrmions. Further, the polarization-reversal-induced change of $d/J$, namely 0.39−0.26=0.13, is also much larger in MnSeTe/In$_2$Se$_3$ than that in the other reported nonpolar-FM/FE heterostructures (Table 1), which is a direct consequence of the strong magnetoelectric coupling. As a result of these large numbers, in the MnSeTe/In$_2$Se$_3$ heterostructure, formation of skyrmions can be achieved by applying an external magnetic field. At the same time, in the presence of a specific magnetic field, FE polarization reversal, which induces a $d/J$ change ±0.13, can change the density and size of the skyrmions, but, depending on the value of the magnetic field, can also eliminate or create skyrmions. The latter can happen because the critical value of the magnetic field to create skyrmions depends on the value of $d/J$.

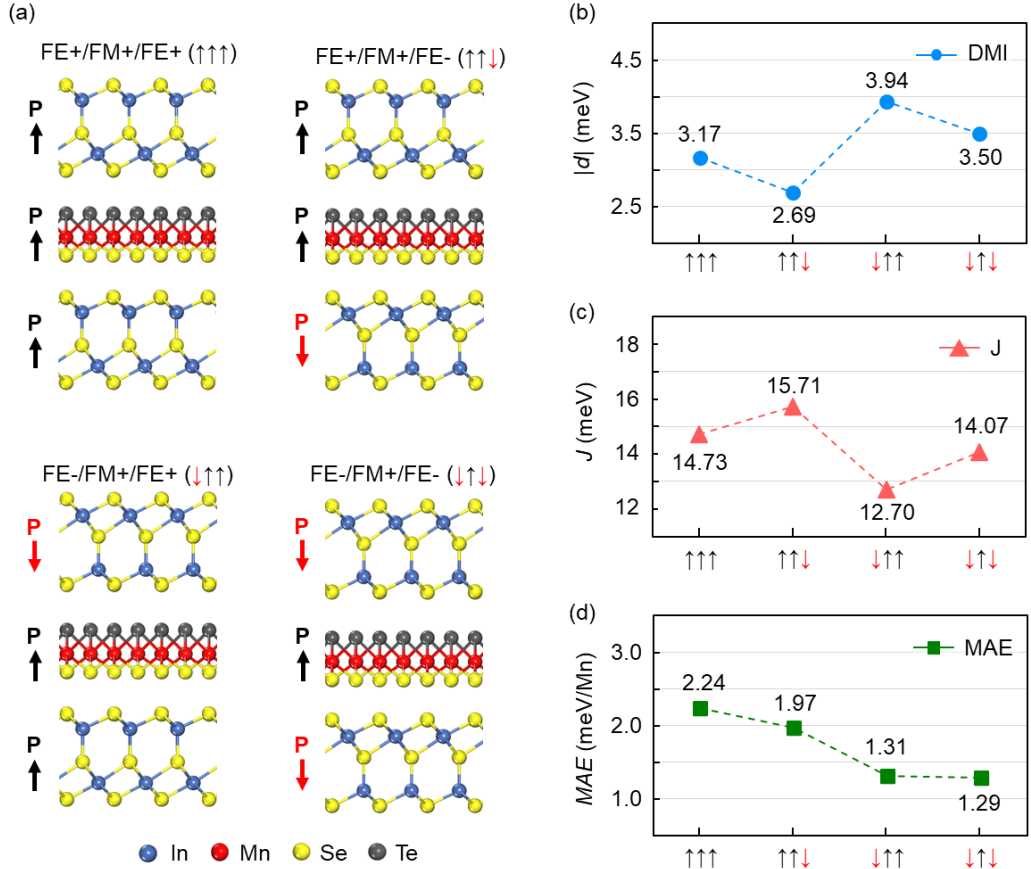

**Figure 2. The modulation of magnetic parameters of MnSeTe in an In$_2$Se$_3$/MnSeTe/In$_2$Se$_3$ heterostructure through the polarization reversal of In$_2$Se$_3$.** (a) Possible configurations of the MnSeTe/In$_2$Se$_3$ heterostructure. (b)-(d) The absolute values of the DMI parameters $|d|$, exchange parameters, $J$ and magnetic-anisotropy energies (*MAE*) of MnSeTe in the four possible heterostructures.

Based on the bilayer MnSeTe/In$_2$Se$_3$ heterostructure, we further built a trilayer heterostructure In$_2$Se$_3$/MnSeTe/In$_2$Se$_3$. Given that MnSeTe in the trilayer heterostructure is metallic, a three-terminal device can be formed by contacting metal electrodes on the surface of the top and bottom In$_2$Se$_3$ layers and a third electrode on the metallic MnSeTe interlayer, which allows multiple configurations of the FE polarizations of the two In$_2$Se$_3$ layers. The several combinations of the polarizations of the top and bottom In$_2$Se$_3$ layers allow tunable and stronger magnetoelectric coupling compared to the bilayer case, as well as multi-state control of skyrmions, as we shall now demonstrate. Intuitively, there are eight different polarization configurations of the trilayer heterostructure, but for every specific polarization combination (e.g., ↑↑↑), there is an equivalent one (e.g., ↓↓↓), finally leading to only four possible configurations, as shown in Fig. 2(a). The $d$, $J$ and $MAE$ of MnSeTe in the four different In$_2$Se$_3$/MnSeTe/In$_2$Se$_3$ heterostructure configurations are shown in Figs. 2(b)-(d), respectively. We list the largest values of $\Delta d$ and $\Delta J$ in Table 1, indicating that the modulation amplitudes of the trilayer heterostructures can be significantly larger than what can be obtained by bilayer heterostructures. In addition, $\Delta d$, $\Delta J$, and $\Delta MAE$ also exhibit different values for different combinations of the FE polarizations, demonstrating the magnetoelectric coupling is also tunable in the trilayer heterostructure. Moreover, as shown in Fig. 2(b) and 2(c), the opposite sings of $\Delta d$ and $\Delta J$ during the polarization-reversal process indicate that magnetic skyrmions can be modulated via multiple polarization states, which is further evidenced by the four different $d/J$ ratios ranging from 0.17 to 0.31, as shown in Table 1.

Finally, we discuss the role of the polarization-polarization (P-P) interactions in the strong magnetoelectric coupling. In the MnSeTe/In$_2$Se$_3$ heterostructure, both MnSeTe and In$_2$Se$_3$ possess intrinsic polarizations. The polarization of MnSeTe is enhanced or suppressed when the FE polarization is reversed due to P-P interactions, which modify the degree of inversion-symmetry breaking in MnSeTe and the strength of associated magnetic interactions (especially DMI). Definitely, such an effect is rather weak in a nonpolar-FM/FE heterostructure. As shown in Table S1, for a MnSeTe/In$_2$Se$_3$ heterostructure, the Mn-Se(Te)-Mn bond angle θ (φ) exhibits a larger change than that of MnSe$_2$/In$_2$Se$_3$ heterostructure when the FE polarization is reversed, validating that P-P interactions enhance the modification of inversion-symmetry breaking. To further check the effect of P-P interactions, we replaced the MnSeTe (In$_2$Se$_3$) in the bilayer heterostructure by a 2D

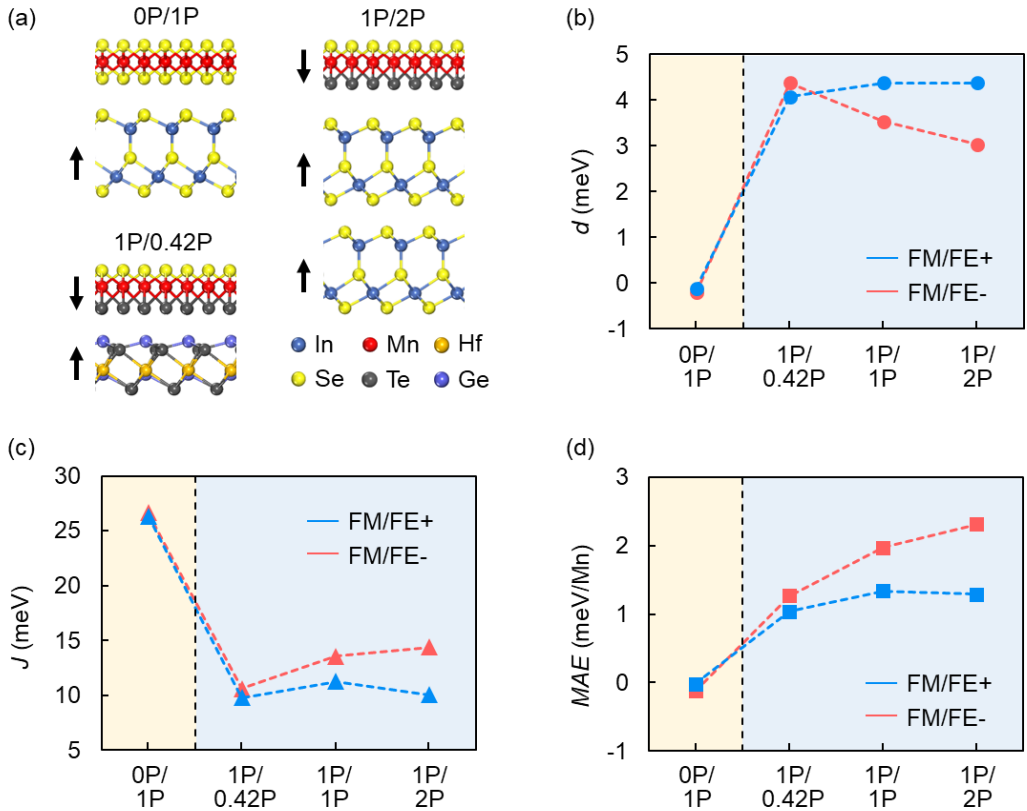

**Figure 3. The effects of polarization-polarization interaction on magnetoelectric coupling.** (a) Atomic configurations of MnSe$_2$/In$_2$Se$_3$, MnSeTe/Hf$_2$Ge$_2$Te$_6$ and MnSeTe/bilayer-In$_2$Se$_3$ heterostructures, which are denoted as 0P/1P, 1P/0.42P and 1P/2P, respectively. Here mP/nP represents the polarizations of FM and FE materials relative to MnSeTe/In$_2$Se$_3$ (1P/1P). (b)-(d) The DMI parameter *d*, exchange parameter *J* and magnetic anisotropic energy (*MAE*) of FM material in the heterostructure.

FM (FE) material with smaller or larger polarizations, whereby the strength of the P-P interactions can be tuned. The obtained systems, MnSe$_2$/In$_2$Se$_3$, MnSeTe/Hf$_2$Ge$_2$Te$_6$, MnSeTe/In$_2$Se$_3$ and MnSeTe/bilayer-In$_2$Se$_3$, are named as 0P/1P, 1P/0.42P, 1P/1P and 1P/2P, respectively, according to the polarizations of FM (FE) materials relative to MnSeTe (In$_2$Se$_3$), as shown in Fig. 3(a). Details of the selected materials are shown in Supplementary Material. The $d$, $J$ and $MAE$ values of these heterostructures are shown in Figs. 3(b)-(d). Evidently, for the heterostructure MnSe$_2$/In$_2$Se$_3$ (0P/1P) without P-P interactions, the ferroelectric-modulation of the magnetic parameters (from red dots to blue dots) is rather small, indicating the magnetoelectric coupling is weak. However, with the enhancement of the P-P interactions (from 1P/0.42P to 1P/2P), the magnetoelectric coupling between FM and FE materials is enhanced significantly. We also note that such a P-P interaction is not a pure electric-field effect, since the changes of magnetic parameters in the MnSeTe/In$_2$Se$_3$ heterostructures are larger than that of MnSeTe under an electric field equivalent to the polarization of In$_2$Se$_3$, as discussed in Fig. S4. Thus, the P-P interaction is the key to achieve the enhanced magnetoelectric coupling.

In this work, we theoretically proposed van der Waals heterostructures composed of a 2D ferromagnetic polar metal, MnSeTe, and a 2D ferroelectric, $In_2Se_3$. The bilayer MnSeTe/$In_2Se_3$ heterostructure exhibits strong magnetoelectric coupling originating from the polarization-polarization interaction between MnSeTe and $In_2Se_3$, resulting in more efficient ferroelectric control of the magnetic interactions and skyrmions in MnSeTe. Furthermore, we proposed a trilayer heterostructure, $In_2Se_3$/MnSeTe/$In_2Se_3$, in which the magnetoelectric coupling is further enhanced and tunable. Such a trilayer heterostructure also allows manipulation of magnetic skyrmions via multiple polarization states. In addition to electrical control of the magnetic skyrmions, the strong magnetoelectric coupling in the MnSeTe/$In_2Se_3$ heterostructure also enables other potential applications. For example, the exchange parameter $J$ exhibits two different values before and after FE polarization reversal, corresponding to two different Curie temperatures (named as T1 and T2 here). Thus, by setting the operating temperature between T1 and T2, a ferromagnetic-to-paramagnetic transition in MnSeTe can be expected by reversing the polarization of $In_2Se_3$. Within mean-field theory, a calculated change of 21% for $J$ in the MnSeTe/$In_2Se_3$ heterostructure would lead to a 21% change in the Curie temperature, which would be sufficient window for a possible application of the effect. Our findings provide insights for achieving strong magnetoelectric coupling in the 2D limit, as well as a new perspective for the design of related spintronics.

**Acknowledgments**: Work in China was supported by the National Natural Science Foundation of China (51922011 and 61888102) and K.C.Wong Education Foundation. Work at Vanderbilt was supported by the U.S. Department of Energy, Office of Science, Division of Basic Energy Sciences, Materials Science and Engineering Directorate grant No. DE-FG-02-09ER46554 and the McMinn Endowment.

**Competing interests**: The authors declare no competing interests.

**Data availability**: Data that support the findings of this study are available from the corresponding authors upon reasonable request.